\documentstyle[twoside,fleqn,psfig,espcrc2]{article}%
\begin{document}                                                        
\renewcommand{\refname}{\normalsize\bf References}
\title{%
Quantum Graphs: A model for Quantum Chaos}

\author{%
        Tsampikos Kottos 
        \address{{Max-Planck-Institut f\"ur Str\"omungsforscung,} \\
{37073 G\"ottingen, Germany }}%
        \,and
        Holger Schanz$^{\rm a,}$%
        \address{{Institut f\"ur Nichtlineare Dynamik, Georg-August-Universit\"at,} \\
{37073 G\"ottingen, Germany }}%
}
%
%
\begin{abstract}
\hrule
\mbox{}\\[-0.2cm]

\noindent{\bf Abstract}\\
We study the statistical properties of the
scattering matrix associated with generic quantum graphs. The scattering
matrix is the quantum analogue of the classical evolution operator on the
graph. For the energy-averaged spectral form factor of the scattering matrix
we have recently derived an exact combinatorial expression. It is based on a
sum over families of periodic orbits which so far could only be performed in
special graphs. Here we present a simple algorithm implementing this summation
for any graph. Our results are in excellent agreement with direct numerical 
simulations for various graphs. Moreover we extend our previous notion of an ensemble of
graphs by considering ensemble averages over random boundary conditions
imposed at the vertices. We show numerically that the corresponding form
factor follows the predictions of random-matrix theory when the number of
vertices is large---even when all bond lengths are degenerate. The
corresponding combinatorial sum has a structure similar to the one obtained
previously by performing an energy average under the assumption of
incommensurate bond lengths.
\\[0.2cm]
\hrule
\end{abstract}

\maketitle
\newcommand{\text}[1]{\mbox{#1}}
\footnotetext[1]{To be published in Physica E, special
issue ''Dynamics in Complex Systems''.}
\section{Introduction}
Quantum graphs have recently attracted a lot of interest
\cite{KS97,SS99,KS99,BG99,ACD+99,BK99,aloc,Tan00}. A review paper containing
a list of references to previous work can be found in \cite{KS97}. The attention
is due to the fact that quantum graphs can be viewed as typical and simple
examples for the large class of systems in which classically chaotic dynamics
implies universal spectral correlations in the semiclassical limit
\cite{BGS84,berry}. Up to now we have only a very limited understanding of the
reasons for this universality. In a semiclassical approach to this problem the
main stumbling block is the intricate interference between the contributions
of (exponentially many) periodic orbits \cite{ADDKKSS93,CPS98}. Using quantum
graphs as model systems it is possible to pinpoint and isolate this central
problem. In graphs, an exact trace formula exists which is based on the
periodic orbits of a mixing classical dynamical system
\cite{KS97,R83}. Moreover the orbits can be specified by a finite symbolic
code with Markovian grammar. Based on these simplifications it is possible to
rewrite the spectral form factor or other two-point correlation functions in
terms of a combinatorial problem \cite{SS99}. This combinatorial problem has
been solved with promising results for selected graphs: It was shown that the
form factor, ensemble averaged over graphs with a single non-trivial vertex
and two attached bonds (2-Hydra) coincides exactly with the random-matrix
result for the $2\times 2$ CUE \cite{SS99}. In \cite{BK99} the short-time
expansion of the form factor for $N$-Hydra graphs (i.~e.\ one central node
with $N$ bonds attached) was computed in the limit $N\to\infty$, and in
\cite{aloc} a periodic-orbit sum was used to prove Anderson localization in an
infinite chain graph with randomized bond lengths. In \cite{Tan00} the form
factor of binary graphs was shown to approach the random-matrix prediction
when the number of vertices increases. In the present paper we develop a
general method to implement the combinatorial sum on a computer.  Our results
are always compared with direct numerical simulations.

Most of the studies on quantum graphs presented up to now were done with a
fixed set of vertex boundary conditions at the vertices. In this case, one of
the main premises to find spectra following the random-matrix predictions is
that {\em all bond lengths of the graph are rationally independent}.  In
contrast, we will consider here also statistical properties of graph spectra
{\em averaged over the set of boundary conditions}. We show that in this case
the form factor is in agreement with the random-matrix theory (RMT)
predictions even when all bond lengths are degenerate. Moreover, we show that
one can express the form factor averaged over the vertex-scattering matrices
as a combinatorial sum over families of orbits. This sum has the same
structure as obtained in \cite{SS99} by performing a spectral average.

This paper is structured in the following way. In the following Section 2, the
main definitions and properties of quantum graphs are given. We concentrate on
the unitary bond-scattering matrix $S_B$ which can be interpreted as a quantum
evolution operator on the graph. Section 3 deals with the corresponding
classical dynamical system. In Section 4, the statistical properties of the
eigenphase spectrum of the bond-scattering matrix $S_B$ are analyzed and
related to the periodic orbits of the classical dynamics.  Finally, our
conclusions are summarized in Section 5.
%
%
\section{Quantum Graphs: Basic Facts}
We start with a presentation and discussion of the Schr\"odinger operator for
graphs.  Graphs consist of $V$ {\it vertices} connected by $B$ {\it bonds}.
The {\it valency} $v_{i}$ of a vertex $i$ is the number of bonds meeting at
that vertex.  The graph is called $v${\it -regular} if all the vertices have
the same valency $v$.  When the vertices $i$ and $j$ are connected, we denote
the connecting bond by $b=(i,j)$.  The same bond can also be referred to as
$\vec{b} \equiv (Min(i,j),Max(i,j))$ or $\smash{\stackrel{\leftarrow}{b}}
\equiv (Max(i,j),Min(i,j))$ whenever we need to assign a direction to the
bond. A bond with coinciding endpoints is called a {\it loop}. Finally, a
graph is called {\em bipartite} if the vertices can be divided into two
disjoint groups such that any vertices belonging to the same group are not
connected.

Associated to every graph is its {\it connectivity (adjacency) matrix}
$C_{i,j}$.  It is a square matrix of size $V$ whose matrix elements $C_{i,j}$
are given in the following way
$$
C_{i,j}=C_{j,i}=\left\{
\begin{array}{l}
1\text{ if }i,j\text{ are connected} \\
0\text{ otherwise}
\end{array}
\right\}\,.  \label{cmat}
$$
For graphs without loops the diagonal elements of $C$ are zero. The
connectivity matrix of connected graphs cannot be written as a block diagonal
matrix. The valency of a vertex is given in terms of the connectivity matrix,
by $v_i= \sum_{j=1}^V C_{i,j}$ and the total number of undirected bonds is $B=
{1\over 2} \sum_{i,j=1}^VC_{i,j}$.

For the quantum description we assign to each bond $b=(i,j)$ a coordinate
$x_{i,j}$ which indicates the position along the bond. $x_{i,j}$ takes the
value $0$ at the vertex $i$ and the value $L_{i,j} \equiv L_{j,i}$ at the
vertex $j$ while $x_{j,i}$ is zero at $j$ and $L_{i,j}$ at $i$. We have thus
defined the {\it length matrix} $L_{i,j}$ with matrix elements different from
zero, whenever $C_{i,j}\neq 0$ and $L_{i,j}=L_{j,i}$ for $b=1,...,B$.  The
wave function $\Psi$ contains $B$ components 
$\Psi_{b_1}(x_{b_1}),\Psi_{b_2}(x_{b_2}),...,\Psi_{b_B}(x_{b_B})$
where the set $\{b_i\}_{i=1}^B$ consists of $B$ different undirected bonds. 

The Schr\"{o}dinger operator (with $\hbar = 2m = 1$) is defined on a graph in
the following way: On each bond $b$, the component $\Psi_b$ of the total wave
function $\Psi$ is a solution of the one-dimensional equation
\begin{equation}
\left( -{\rm i}{\frac{{\rm d\ \ }}{{\rm d}x}}-A_b\right) ^2\Psi _b(x)=k^2\Psi
_b(x)\,.\label{schrodinger}
\end{equation}
We included a ``magnetic vector potential" $A_b$ (with $\Re e(A_{b})\ne 0$
and $A_{\vec{b}}= -A_{_{\smash{\stackrel{\leftarrow}{b}}}}$) which breaks
the time reversal symmetry. In most applications we shall assume that all
the $A_{b}$'s are equal and the bond index will be dropped.
On each of the bonds, the general solution of (\ref{schrodinger}) is a
superposition of two counter propagating waves
\begin {equation}
\Psi_{b=(i,j)} = a_{i,j} {\rm e}^{{\rm i}(k+A_{i,j})x_{i,j}}
+a_{j,i} {\rm e}^{{\rm i}(k+A_{j,i})x_{j,i}}
\label{propa}
\end{equation}
The coefficients $a_{i,j}$ form a vector ${\bf a}\equiv (a_{\vec{b}_1}$,
$\cdots$, $ a_{\vec{b}_B},$
$a_{_{\smash{\stackrel{\leftarrow}{b}_1}}}, \cdots, 
a_{_{\smash{\stackrel{\leftarrow}{b}_B}}})^T$ of complex numbers which
uniquely determines an element in a $2B-$dimensional Hilbert space. This space
corresponds to "free wave" solutions since we did not yet impose any
conditions which the solutions of (\ref{schrodinger}) have to satisfy at the
vertices.

The most general boundary conditions at the vertices are given in terms of
unitary $v_j\times v_j$ {\it vertex-scattering matrices}
$\sigma^{(j)}_{l,m}(k)$, where $l$ and $m$ go over all the vertices which are
connected to $j$.  At each vertex $j$, incoming and outgoing components of the
wave function are related by
\begin{equation}
a_{j,l}=\sum_{m=1}^{v_j}\sigma _{l,m}^{(j)}(k)e^{{\rm i}kL_{jm}} a_{m,j} \,,
\label{outin1}
\end{equation}
which implies current conservation. The particular form
\begin{equation}\label{Neumann}
\sigma^{(j)}_{l,m}={2\over v_{j}}-\delta_{l,m}
\end{equation}
for the vertex-scattering matrices was shown in \cite{KS97} to be compatible
with continuity of the wave function and current conservation at the vertices.
(\ref{Neumann}) is referred to as {\em Neumann} boundary conditions.

Stationary states of the graph satisfy (\ref{outin1}) at each vertex.  These
conditions can be combined into
\begin{equation}
{\bf a}= S_{B}(k)\, {\bf a}\,,
\label{outin2}
\end{equation}
such that the secular equation determining the eigenenergies and the
corresponding eigenfunctions of the graph is of the form \cite{KS97}
\begin{equation}
\det \left[ I-S_B(k,A)\right] =0\,.  \label{secular}
\end{equation}
Here, the unitary {\em bond-scattering matrix}
\begin{equation}\label{sb-matrix}
S_B(k,A)=D(k;A)\,T 
\end{equation}
acting in the $2B$-dimensional space of directed bonds has been
introduced. The matrices $D$ and $T$ are given by
\begin{eqnarray}
\label{DandT}
D_{ij,i^{\prime }j^{\prime }}(k,A) &=&\delta _{i,i^{\prime }}\delta
_{j,j^{\prime }}{\rm e}^{{\rm i}kL_{ij}+{\rm i}A_{i,j}L_{ij}}\ ;\label{scaco} \\
\ T_{ji,nm} &=&\delta _{n,i}C_{j,i}C_{i,m}\sigma _{j,m}^{(i)}\,. \nonumber
\end{eqnarray}
$T$ contains the topology of
the graph and is equivalent to the complete set of vertex-scattering matrices,
while $D$ contains the metric information about the bonds.

It is instructive to interpret the action of $S_B$ on an arbitrary graph state
$\Psi$ as its time evolution over an interval corresponding to the mean bond
length of the graph such that
\begin{equation}
\label{qevol}
{\bf a}(n) = S_B^n\,{\bf a}(0),\,\,\,\,n=0,1,2,...\,.
\end{equation}
Clearly the solutions of (\ref{outin2}) are stationary with respect to this
time evolution. $n$ in (\ref{qevol}) represents a discrete (topological) time
counting the collisions of the particle with vertices of the graph.

\section{Periodic orbits and classical dynamics on graphs}

In this section we discuss the classical dynamics corresponding to the quantum
evolution (\ref{qevol}) implied by $S_{B}$. To introduce this dynamics we
employ a Liouvillian approach, where a classical evolution operator assigns
transition probabilities in a phase space of $2B$ directed bonds
\cite{KS97}. If $\rho _b(t)$ denotes the probability to occupy the (directed)
bond $b$ at the (discrete) topological time $t$, we can write down a Markovian
Master equation of the form
\begin{equation}
\rho _b(t+1)=\sum_{b^{\prime }}U_{b,b^{\prime }}\rho _{b^{\prime }}(t)\,.
\label{master}
\end{equation}
The classical (Frobenius-Perron) evolution operator $U$ has matrix elements
\begin{equation}
U_{ij,nm}=\delta_{j,n} P^{(j)}_{i\rightarrow m}
\label{cl3}
\end{equation}
with $P_{ji\rightarrow ij^{\prime }}^{(i)}$ denoting the transition
probability between the directed bonds $b=(j,i)$ and $b^{\prime }=(i,j^{\prime
})$. To make the connection with the quantum description, we adopt the quantum
transition probabilities, expressed as the absolute squares of matrix elements
of $S_B$
\begin{equation}
P_{j\rightarrow j^{\prime }}^{(i)}=\left| \sigma_{j,j^{\prime
}}^{(i)}(k)\right| ^2\,.  \label{cl1}
\end{equation}
Note that $P_{j\rightarrow j^{\prime }}^{(i)}$ and $U$ do not involve any
metric information on the graph. In general, they may depend on the wave
number $k$. 

The unitarity of the bond-scattering matrix $S_B$ guarantees
$\sum_{b=1}^{2B}U_{b,b^{\prime }}=1$ and $0\leq U_{b,b^{\prime }}\leq 1$, so
that the total probability that the particle is on any bond remains conserved
during the evolution. The spectrum of $U$ is restricted to the interior of the
unit circle and $\nu_1 = 1$ is always an eigenvalue with the corresponding
eigenvector $|1\rangle = \frac 1{2B} \left( 1,1,...,1\right) ^T$.  In most
cases, the eigenvalue $1$ is the only eigenvalue on the unit circle.  Then,
the evolution is ergodic since any initial density will evolve to the
eigenvector $|1\rangle $ which corresponds to a uniform distribution
(equilibrium). The rate at which equilibrium is approached is determined by
the gap to the next largest eigenvalue. If this gap exists, the dynamics is
also mixing. Graphs are one dimensional and the motion on the bonds is simple
and stable. Ergodic (mixing) dynamics is generated because at each vertex a
(Markovian) choice of one out of $v$ directions is made. Thus, chaos on graphs
originates from the multiple connectivity of the (otherwise linear) system
\cite{KS97}.

Despite the probabilistic nature of the classical dynamics, the concept of a
classical orbit can be introduced.  A classical orbit on a graph is an
itinerary of successively connected directed bonds $(i_1,i_2),
(i_2,i_3),\cdots $. An orbit is {\it periodic} with period $n$ if for all $k$,
$(i_{n+k},i_{n+k+1}) = (i_k,i_{k+1})$. For graphs without loops or multiple
bonds, the sequence of vertices $i_1,i_2, \cdots$ with $i_m \in [1,V]$ and
$C_{i_m,i_{m+1}} =1$ for all $m$ represents a unique code for the orbit. This
is a finite coding which is governed by a Markovian grammar provided by the
connectivity matrix.  In this sense, the symbolic dynamics on the graph is
Bernoulli. This analogy is strengthened by further evidence: The number of
$n-$PO's on the graph is ${\frac 1n}{\rm tr}C^n$, where $C$ is the
connectivity matrix. Since its largest eigenvalue $\Gamma_c$ is bounded
between the minimum and the maximum valency i.e.  $\min v_i \leq \Gamma_c \leq
\max v_i$, periodic orbits proliferate exponentially with topological entropy
$\approx \log \Gamma_c$.


\section {The spectral statistics of $S_B$}
We consider the matrix $S_B(k,A)$ defined in (\ref{sb-matrix}), (\ref{scaco}).
The spectrum consist of $2B$ points $e^{{\rm i}\theta_l(k)}$ confined to the
unit circle (eigenphases).  Unitary matrices of this type are frequently
studied since they are the quantum analogues of classical, area preserving
maps. Their spectral fluctuations depend on the nature of the underlying
classical dynamics \cite{S89}.  The quantum analogues of classically
integrable maps display Poissonian statistics while in the opposite case of
classically chaotic maps, the eigenphase statistics conform with the results
of RMT for Dyson's {\it circular ensembles}. To describe the spectral
fluctuations of $S_{B}$ we consider the form factor
\begin{equation}\label{ff} 
K(n/2B)={\frac 1{2B}}\left\langle |{\rm tr}S_B^n|^2\right\rangle 
\qquad(n>0)\,.
\end{equation} 
The average $\langle\dots\rangle$ will be specified below. 
RMT predicts that $K(n,2B)$ depends on the scaled time $\tau=\frac {n}{2B}$
only \cite{S89}, and explicit expressions for the orthogonal and the unitary
circular ensembles are known \cite{M90}. 

Using (\ref{sb-matrix}), (\ref{DandT}) we expand the matrix products in ${\rm
tr}S_{B}^{n}$ and obtain a sum of the form
\begin{equation}
{\rm tr}S^n_B(k)
=\sum_{p\in{\cal P}_n}{\cal A}_p{\rm e}^{{{\rm i} (kL_p+Al_p)}}\,.
\label{posum}
\end{equation}
In this sum $p$ runs over all closed trajectories on the graph which are
compatible with the connectivity matrix and which have the topological length
$n$, i.~e. they visit exactly $n$ vertices. For graphs, the concepts of closed
trajectories and periodic orbits coincide, hence (\ref{posum}) can also be
interpreted as a periodic-orbit sum. From (\ref{posum}) it is clear that
$K(n/2B)=0$ as long as $n$ is smaller than the period of the shortest periodic
orbit.  The phase associated with an orbit is determined by its total (metric)
length $L_p = \sum_{b \in p} L_{b}$ and by the ``magnetic flux" through the
orbit.  The latter is given in terms of its total {\em directed} length
$l_{p}$ if we assume for simplicity that the magnitude of the magnetic vector
potential is constant $|A_{b}|\equiv A$.  The complex amplitude of the
contribution from a periodic orbit by the product of all the elements of
vertex-scattering matrices encountered
\begin{equation}
{\cal A}_p=\prod_{j=1}^{n_p}\sigma^{(i_j)}_{i_{j-1},i_{j+1}}
\equiv \prod_{[r,s,t]} \left(\sigma^{(s)}_{r,t}\right)^{n_p(r,s,t)}\,,
\end{equation}
i.~e.\ for fixed boundary conditions at the vertices it is completely
specified by the frequencies $n_p(r,s,t)$ of all transitions $(r,s)\to(s,t)$ .
Inserting (\ref{posum}) into the definition of the form factor we obtain a
double sum over periodic orbits
\begin{eqnarray}\label{double-sum}
&&K(n/2B)=\frac 1{2B}\, \Big\langle\sum_{p,p'\in {\cal P}_n} {\cal A}_p{\cal
A}_{p\prime}^{*}\,\times \nonumber\\
&&\quad\exp \left\{{\rm i} k(L_{p}-L_{p'})+{\rm i}
A(l_p-l_{p\prime})\right\}\Big\rangle\,. 
\end{eqnarray}
This will be our starting point for the combinatorial approach presented in the
following subsection.

\subsection{Energy Average}
In previous work \cite{SS99,BK99,aloc} the average in (\ref{double-sum}) was
always taken with respect to the wave number $k$ and, if present, also over
the magnetic vector potential $A$. Within the present subsection we will stick
to this procedure.  Provided that the bond lengths of the graph are rationally
independent and that a sufficiently large interval is used for averaging, only
terms with $L_{p}=L_{p'}$ and $l_{p}=l_{p'}$ survive. A completely equivalent
result can be obtained by regarding the bond lengths $L_{b}$ of the graph as
random numbers and performing an ensemble average over them.
\begin{figure}
\centerline{\psfig{figure=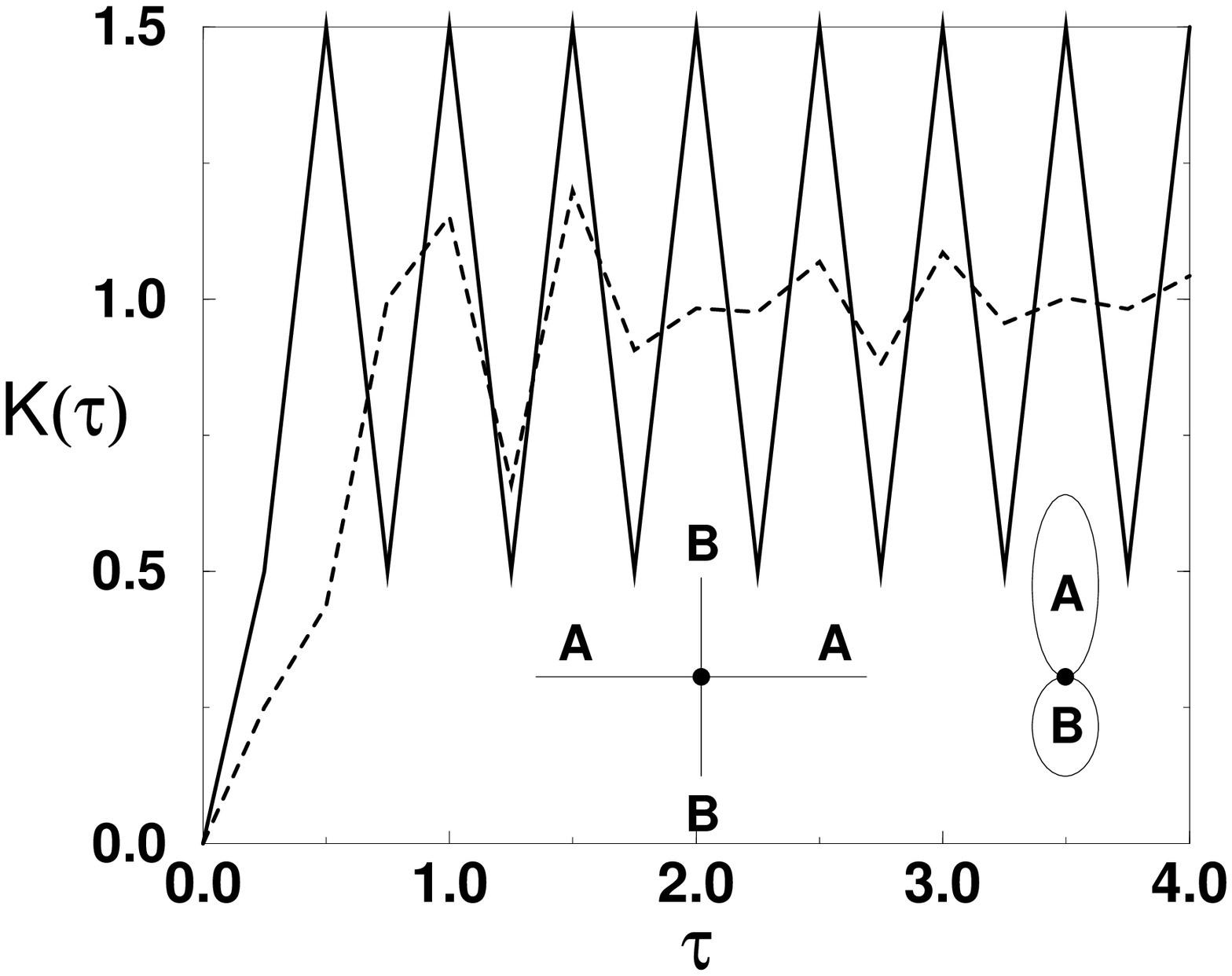,width=60mm}}
\noindent\footnotesize{
{\bf Fig.~1}: Form factor of $S_{B}$ for a double-loop graph (or
equivalently $S_{B}^{2}$ for a 4-Hydra). The full line shows
(\ref{double-loop}) corresponding to zero magnetic field in the loops or two
pairs of degenerate bond lengths in the Hydra as shown in the inset.  Dashed
line (obtained from (\ref{coeffs})): A magnetic flux threads the loops and is
averaged over. This corresponds to 4 incommensurate bond lengths of the Hydra
graph.}
\end{figure}

Note that $L_{p}=L_{p'}$ does {\em not} necessarily imply $p=p'$ or that
$p,p'$ are related by some symmetry because there exist families $\cal L$ of
distinct but isometric orbits which can be used to write the result of
(\ref{double-sum}) in the form \cite{SS99,BK99,aloc}
\begin{equation}\label{family-sum}
K(n/2B)=\sum_{{\cal L}\in{{\cal F}_{n}}}\left|\sum_{p\in{\cal L}}{\cal A}_{p}\right|^2\,.
\end{equation}
The outer sum is over the set ${\cal F}_{n}$ of families, while the inner one
is a {\it coherent} sum over the orbits belonging to a given family (= metric
length). (\ref{family-sum}) is exact, and it represents a combinatorial
problem since it does not depend anymore on metric information about the graph
(the bond lengths).

Let us illustrate the application of (\ref{family-sum}) in the following simple 
system: We consider a Hydra graph with
four arms and impose Neumann boundary conditions at the central node such that
forward and backward scattering amplitudes are given according to (\ref{Neumann})
by $\sigma_{f}=1/2$ and $\sigma_{b}=-{1/2}$. For the moment we require
that there are two pairs of identical bond lengths $L_{A}, L_{B}$ (see
Fig.~1). Since Hydra is a bipartite graph, an orbit can return only
after an even number of vertex-scattering events (note that the dead ends of
the arms are considered as vertices). It is then convenient to replace the
scattering matrix in (\ref{ff}) by $S_{B}^{2}$. As one can see from 
(\ref{posum}), that the form factor obtained in this way
is equivalent to the one obtained from a graph which is formed by two loops of
lengths $2L_{A}$ and $2L_{B}$ connected by a single vertex (inset of
Fig.~1). Consider now the periodic orbits contributing to $K(n/2B)$. The
length of such an orbit is given by $L=n_{A}L_{A}+n_{B}L_{B}$ where $n_{A}$,
$n_{B}$ denotes the number of traversals of A and B loop, respectively, and
$n=n_{A}+n_{B}$. If $L_{A}, L_{B}$ are rationally independent, a family of
isometric orbits in (\ref{family-sum}) contains all orbits sharing $n_{A}$,
$n_{B}$. These orbits differ, however, in their symbolic code by the order of
A's and B's and by the {\em orientation} in which the loops are
traversed. This is reflected in the amplitudes ${\cal A}_{p}$ of the orbits:
If two consecutive loops AA have different orientation, the amplitude ${\cal
A}_{p}$ contains a factor $\sigma_{b}$, while the same orientation results in
$\sigma_{f}$. Different loops following each other (AB or BA) always result in
a factor $\sigma_{f}$. Suppose now there is an orbit which contains in its
code a sequence of the form $\cdots$AB$_1$B$_2$B$_s\cdots A\cdots$. Then there
will be another orbit $\cdots$AB$_1$B'$_2$B'$_s\cdots A\cdots$ which differs
from the first one by an inversion of the orientation (denoted by the prime)
of all but the first loop B. It is easy to see that the corresponding
amplitudes cancel. This results in a tremendous simplification of the problem:
It suffices in (\ref{family-sum}) to keep track of all trajectories for which
the cancellation does not apply. These are of the form A$^{n}$, B$^{n}$,
(AB)$^{n/2}$ or (BA)$^{n/2}$, the latter two existing for even $n$ only.  Each
orbit A$^{n}$ contributes the amplitude $1/2^{n}$ irrespective of the
orientations of the $n$ loops. On the other hand there are $2^{n}$
possibilities to prescribe these orientations and hence the total contribution
from each of the families A$^{n}$ and B$^{n}$ is 1. There are also $2^{n}$
trajectories of the form (AB)$^{n/2}$ and (BA)$^{n/2}$, respectively, each
with amplitude $1/2^{n}$. However these orbits belong to the same family
$n_{A}=n_{B}=n/2$, and their amplitudes must be added coherently in
(\ref{family-sum}). This results in the total contribution 4 from this
family. After adding the contributions from the three different families and
normalizing with $2B=4$ we are left with
\begin{equation}\label{double-loop}
K(n/2B)=\left\{\begin{array}{ll}1/2&n=1,3,\dots\\ 3/2&n=2,4,\dots\end{array}\right.\,,
\end{equation}
which is shown in Fig.~1 with a solid line. Neglecting the point $n=0$, the
form factor simply oscillates around the value 1---even within the Heisenberg
time $\tau=n/2B=1$. This is reminiscent of the behavior of a Poissonian
spectrum. The reason for this lack of correlations is that there exist regular
sequences of eigenstates which are completely confined to one of the two loops
or one pair of degenerate bonds, respectively. These states disappear when all
arms of the Hydra are incommensurate, and the corresponding form factor
(dashed line in Fig.~1) clearly reflects this change in the properties of the
spectrum.

The arguments which led to (\ref{double-loop}) are deceivingly simple.  In
general, the combinatorial problem (\ref{family-sum}) is very hard and cannot
be solved in closed form. Even for the 4-Hydra with incommensurate bond
lengths this seems to be the case. Nevertheless {\em exact} result for finite
$n$ as those shown in Fig.~1 with a dashed line can always be obtained from
(\ref{family-sum}) using a computer algebra system such as Maple \cite{Maple}.
This will be shown in the following.  To prescribe topology and boundary
conditions, one should provide the time evolution operator $S_{B}$ according
to (\ref{sb-matrix}) but with the phases ${\rm e}^{{\rm i} kL_{i,j}}$ left as
unspecified variables. For example, the operator $S_{B}$ for the graph
discussed above has the form 
{\small
$$
\left [\begin {array}{cccc} 
-1/2\,\phi_{{A}}&1/2\,\phi_{{A}}&1/2\,\phi_{{B}}&1/2\,\phi_{{B}}
\\\noalign{\medskip}
1/2\,\phi_{{A}}&-1/2\,\phi_{{A}}&1/2\,\phi_{{B}}&1/2\,\phi_{{B}}
\\\noalign{\medskip}
1/2\,\phi_{{A}}&1/2\,\phi_{{A}}&-1/2\,\phi_{{B}}&1/2\,\phi_{{B}}
\\\noalign{\medskip}
1/2\,\phi_{{A}}&1/2\,\phi_{{A}}&1/2\,\phi_{{B}}&-1/2\,\phi_{{B}}
\end {array}\right ],
$$
} where $\phi_{A/B}={\rm e}^{{\rm i} 2kL_{A/B}}$. In the general case there will be as
many phases $\phi_{i}$ as there are incommensurate lengths in the graph. The
quantity ${\rm tr} S_{B}^{n}$ can now be represented as a multivariate polynomial
of degree $n$ in the variables $\phi_{i}$, i.~e.\
\begin{equation}
{\rm tr} S_{B}^{n}=\sum_{{\cal P}_{n}}c_{\cal P}\,\phi_{1}^{p_{1}}\,\phi_{2}^{p_{2}}\dots\,,
\end{equation}
where ${\cal P}_{n}$ runs over all partitions of $n$ into non-negative
integers $n=p_{1}+p_{2}+\dots$. The form factor is then simply given as 
\begin{equation}\label{coeffs}
K(n/2B)=\sum_{{\cal P}_{n}}|c_{\cal P}|^{2}\,.
\end{equation}
The task of finding the coefficients $c_{\cal P}$ can be expressed in Maple
with standard functions, such that after the initialization of $S_{B}$ the
following four simple lines represent a completely  general algorithm for the 
exact computation of the form factor of an arbitrary graph:
{\small
\begin{verbatim}
  sn:=evalm(S^n);
  tn:=expand(trace(sn)):
  c:=[coeffs(tn)];
  K:=sum(c[p]^2,p=1..nops(c));
\end{verbatim}
} In practice, however, the computation is restricted to the first few $n$
since the numerical effort grows exponentially fast. In Fig.~2 we compare the
results of (\ref{coeffs}) with direct numerical averages for regular (fully
connected) graphs with $V=4$ and $V=5$ vertices with and without magnetic
field breaking the time-reversal symmetry. The results agree indeed to a high
precision. Although this could be regarded merely as an additional
confirmation of the numerical procedures used in \cite{KS97}, we see the
main merit of (\ref{coeffs}) in being a very useful tool for trying to find
the solution of (\ref{family-sum}) in closed form.
\begin{figure}
\vspace*{-0cm}
\centerline{\psfig{figure=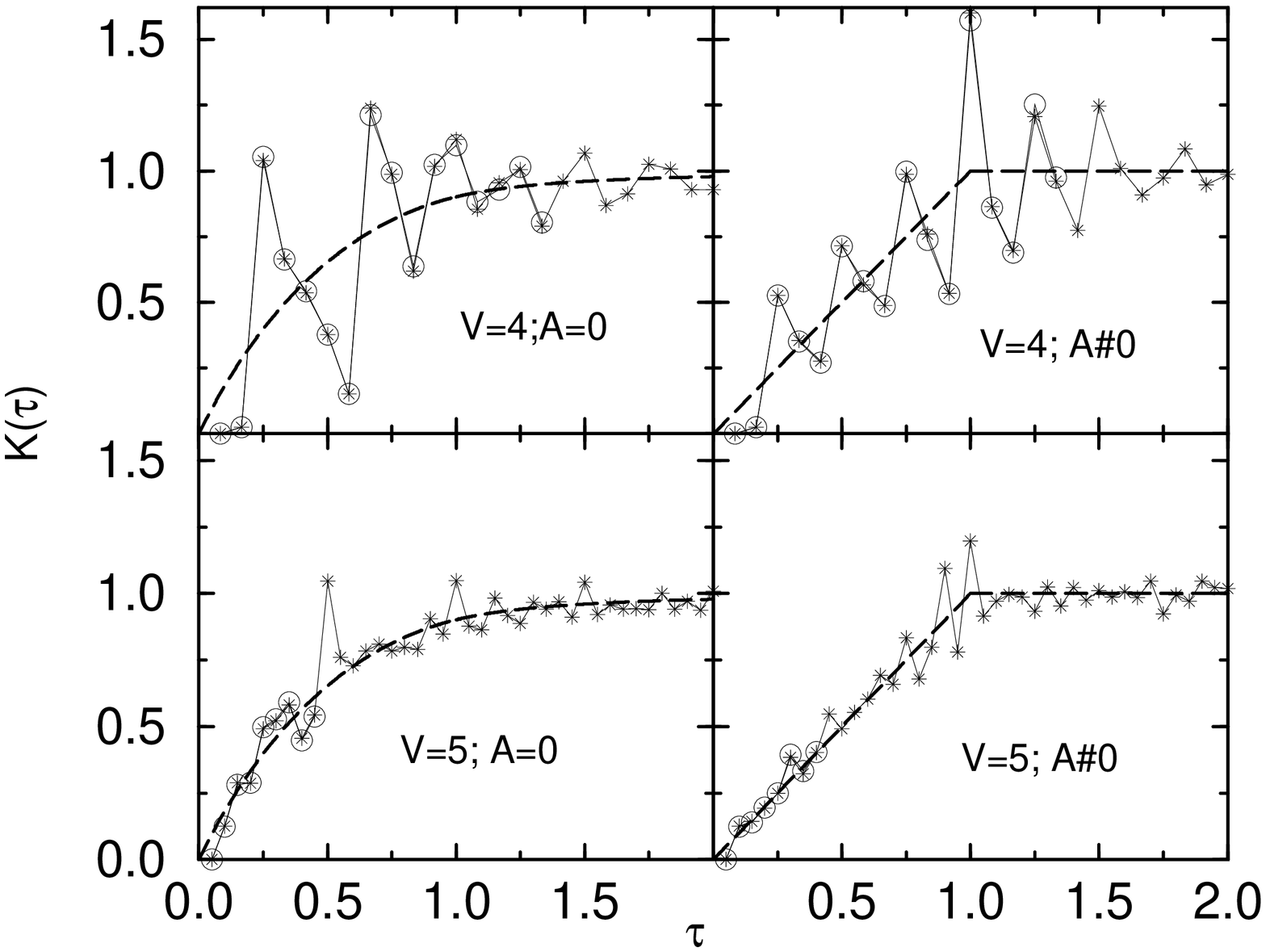,width=75mm,angle=270}}
\noindent
\footnotesize {{\bf FIG. 2.}
Form factor of $S_{B}$ for regular graphs with $V=4$
vertices (top) and $V=5$ vertices (bottom). In the right panels an additional
magnetic field destroyed time-reversal symmetry. Circles: exact results
obtained from (\protect\ref{family-sum}). Stars: numerical average over 2,000
values of $k$. The solid line is to guide the eye. The prediction of the
appropriate random matrix ensemble are shown with
dashed lines.}
\end{figure}
\subsection{Ensemble Average over the Boundary Conditions at Vertices}
In the present section we generate an ensemble of graphs by randomizing the
vertex-scattering matrices $\sigma^{(i)}$. The length matrix $L$ which
contains the lengths of the bonds and the connectivity matrix (topology of the
graph) are kept constant. Moreover all the bond lengths are equal $L_{b}=1$. We
computed the form factor for fully connected graphs. Our results for values of
$V=10,15,20$ are presented in Fig.~3. The RMT two-point form factor \cite{M90}
is also displayed in Fig.~3 for comparison. The results show quite a good
agreement with the predictions of RMT for the circular ensembles. Moreover,
one can see that as the number of vertices $V$ increases, the agreement with
RMT becomes progressively better. The deviations from the smooth curves are not
statistical, and cannot be ironed out by further averaging. Rather, they are
due to the fact that the graph is a dynamical system which cannot be described
by RMT in all detail.
\begin{figure}
\vspace*{-0mm}
\centerline{\psfig{figure=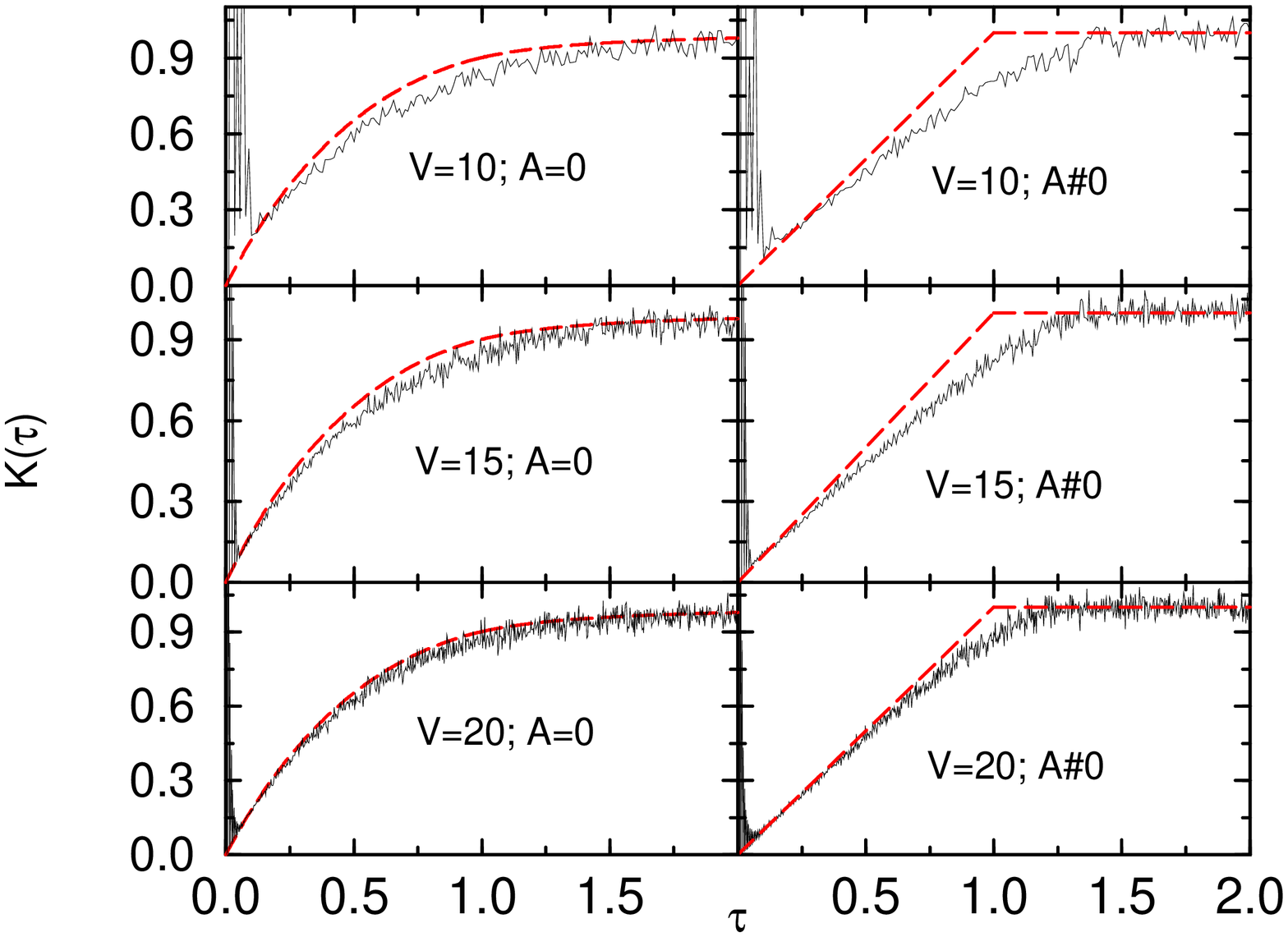,width=75mm,angle=270}}
\noindent\footnotesize{ 
{\bf Fig.~3:} The form factor of the eigenphase spectrum of $S_B$ for fully
connected graphs with $V=10,15,20$. Bold dashed lines are the expectations for
the COE and CUE expressions. The data are averaged over realizations of the
$\sigma$'s, as explained in the text. All the bond lengths are kept constant and
equal to $L_{i.j}=1$.}
\end{figure}

In the remainder of this subsection, we would like to point out the
possibility for turning (\ref{double-sum}) into a combinatorial problem using
the ensemble-average over $\sigma^{(i)}$.  We assume that $\sigma^{(i)}$
$i=1,\dots V$ are independent random matrices chosen from the CUE. In this
case, the averaging in (\ref{double-sum}) factorizes with respect to the
vertices and we find
\begin{eqnarray}\label{orbit-pair}
\left\langle {\cal A}_p{\cal A}_{p\prime}^{*}\right\rangle
&=&\prod_{s}\Big\langle \prod_{[r,t]}
\left(\sigma^{(s)}_{r,t}\right)^{n_{p}(r,s,t)} \times\nonumber\\
&&
\qquad\left(\sigma^{(s)*}_{r,t}\right)^{n_{p'}(r,s,t)}
\Big\rangle 
\end{eqnarray}
Averages over the CUE of this type have been computed in the literature, see
e.~g.\ \cite{BB96}. The result is of the form
\begin{eqnarray}
&&\left\langle U_{a_{1},b_{1}}\dots U_{a_{n},b_{n}}\,\times\,U_{a_{1},b_{1}}\dots 
U_{\alpha_{n},\beta_{n}}\right\rangle_{CUE}= \nonumber\\
&&\qquad\sum_{P,P'}C_{P^{-1}P'}\prod_{j=1}^{n}\delta_{a_{j},\alpha_{P(j)}}
\,\delta_{b_{j},\beta_{P'(j)}}\,,
\end{eqnarray}
where $P,P'$ run over the permutations of $1,\dots,n$ and $C_{P^{-1}P'}$ are
some constants which can be obtained from a recursion relation. Applied to
(\ref{orbit-pair}) this implies, that the contribution from a pair of orbits
$p,p'$ vanishes unless for all $s,t$
$\sum_{r}n_{p}(r,s,t)=\sum_{r}n_{p'}(r,s,t)$
$\sum_{r}n_{p}(r,s,t)=\sum_{r}n_{p'}(r,s,t)$, i.~e.\ unless the frequencies of
tra\-versals of any directed bond $s\to t$ coincide for $p$ and $p'$.  For the
case of a graph with time-reversal symmetry, i.~e.\ when the vertex-scattering
matrices are taken at random from the COE, a similar result ensures that only
those pairs of orbits survive the averaging in (\ref{orbit-pair}) which agree
in the traversals of all {\em undirected} bonds.  This means that the family
structure underlying the set of periodic orbits is exactly the same as in the
case of a graph with incommensurate bond lengths but fixed boundary
conditions: The families are formed by those orbits which differ only in the
time order of the traversed bonds.  Substituting this result into (\ref{double-sum})
we see that all oscillating phases containing metric information on the graph
drop independent of the precise values of the bond lengths and we are indeed left
with a combinatorial problem.

\section{Conclusions}

In this paper we have tried to contribute to the understanding of the
statistical properties of the unitary quantum time evolution operator derived
from quantum graphs. This problem is relevant since it is a paradigm for the
as yet unanswered question precisely under which conditions the quantum
analogues of classically chaotic systems are universal and follow the
random-matrix predictions. Fully connected quantum graphs show this
universality when the number of bonds becomes large. In extension to previous
work we have demonstrated this result in the case where an ensemble of graphs
is introduced by randomizing the boundary conditions at the vertices. The
corresponding ensemble average can replace the previously considered spectral
average, and the RMT results are in this case approached even when all the
bond lengths are incommensurate. We have the hope that this kind of ensemble
average might provide an easier access to an analytical treatment of the
spectral properties of graphs. One possible approach of this goal is the use
of combinatorial methods to perform the periodic-orbit sums related to
spectral two-point correlations. 

\section*{Acknowledgements}
We would like to express our gratitude to Prof.~Uzy Smilansky to whom we owe
our interest in the subject and who contributed a great deal to most results
discussed in this paper.


\end{document}